\documentstyle[prl,aps,preprint]{revtex}
\newcommand{\tr}{{\rm tr}}
\newcommand{\Tr}{{\rm Tr}}
\newcommand{\D}{{\rm D}}
\newcommand{\Rc}{{R_c}}
\newcommand{\thru}[1]{\mathrel{\mathop{#1\!\!\!/}}}

\newcommand{\bfd}{\mbox{\boldmath $d$}}
\newcommand{\bfk}{\mbox{\boldmath $k$}}
\newcommand{\bfp}{\mbox{\boldmath $p$}}
\newcommand{\bfv}{\mbox{\boldmath $v$}}
\newcommand{\bfx}{\mbox{\boldmath $x$}}
\newcommand{\bfA}{\mbox{\boldmath $A$}}
\newcommand{\bfD}{\mbox{\boldmath $D$}}
\newcommand{\bfF}{\mbox{\boldmath $F$}}

\newcommand{\bfV}{\mbox{\boldmath $V$}}
\newcommand{\bfgamma}{\mbox{\boldmath $\gamma$}}
\newcommand{\Ar}{{\cal A}}
\newcommand{\Dr}{{\cal D}}
\newcommand{\Fr}{{\cal F}}
\newcommand{\Vr}{{\cal V}}
\newcommand{\Ur}{{U}}
\newcommand{\bfAr}{\mbox{\boldmath ${\cal A}$}}
\newcommand{\bfFr}{\mbox{\boldmath ${\cal F}$}}
\newcommand{\bfVr}{\mbox{\boldmath ${\cal V}$}}

\begin{document}

\draft
\tighten
\def\footnoterule{\kern-3pt \hrule width\hsize \kern3pt}

\title{
Temperature dependence of the anomalous effective 
action of fermions in two and four dimensions
}

\author{L.L. Salcedo}

\address{
{~} \\
Departamento de F\'{\i}sica Moderna \\
Universidad de Granada \\
E-18071 Granada, Spain
}

\date{\today}
\maketitle

\thispagestyle{empty}

\begin{abstract}
The temperature dependence of the anomalous sector of the effective
action of fermions coupled to external gauge and pseudo-scalar fields
is computed at leading order in an expansion in the number of Lorentz
indices in two and four dimensions. The calculation preserves chiral
symmetry and confirms that a temperature dependence is compatible with
axial anomaly saturation. The result checks soft-pions theorems at
zero temperature as well as recent results in the literature for the
pionic decay amplitude into static photons in the chirally symmetric
phase. The case of chiral fermions is also considered.
\end{abstract}



\section{Introduction}\label{sec:1}

It is well established in the literature that the axial anomaly has a
temperature independent form
~\cite{Do74,It83,Re85,Ru86,Da87,Co88,Li88,Wa90,He90,Ba91,Ra92,Sm92,Go94}.
This result is consistent with our present understanding of anomalies,
since they are induced by the ultraviolet divergences present in the
theory whereas the finite temperature modifies the infrared sector
only, namely, by imposing periodic or antiperiodic boundary conditions
in the Euclidean time direction~\cite{La87,Ka89,Be96}. In particular
for Weyl fermions, the known topological origin of the anomaly
guarantees its independence under changes on the space-time
manifold~\cite{Al85}. Likewise, the axial anomaly is mass independent
and also density independent~\cite{Go94}. Analogous statements hold
for the parity anomaly in odd-dimensional theories~\cite{Sa98}.

Recently, it has been found in~\cite{Pi96} that the anomalous
amplitudes, such as the neutral pion decay into two photons, are
temperature dependent. At first, this would seem surprising since the
numerical value of the pion width is very well accounted for by the
axial anomaly prediction and furthermore, such decay is considered the
standard proof that chiral anomalies are not just mathematical
artifacts. In order to clarify this point it should be noted that even
at zero temperature the axial anomaly and the neutral pion decay are
different entities; Adler's theorem~\cite{Ad69}, relates the pion
decay to the axial anomaly in the soft-pion limit only. Translated to
the language of effective actions, this means that, at leading order
in a gradient expansion, the pseudo-parity odd component of the
effective action~\cite{Wi83} (or equivalently, the component that is
imaginary in Euclidean time or also the component containing the
Levi-Civita pseudo-tensor) is accounted for by the gauged
Wess-Zumino-Witten (WZW) action~\cite{We71,Wi83}. In other words, all
other contributions are of higher order. They contain more gradients
and are therefore subdominant in the soft-pion limit. Such higher
order terms will necessarily be chirally invariant since the gauged
WZW action saturates the anomaly. If one tries to write down Euclidean
(Lorentz) and chiral invariant terms of the same order as the WZW
action, it is immediately clear that they vanish identically. At
finite temperature the conditions are less restrictive and the pion
decay is no longer determined by the axial anomaly~\cite{Pi97a,It83}.
Indeed, at finite temperature the time direction is privileged and
Euclidean invariance is partially broken. Moreover, the effective
action is not expected to admit a gradient expansion with local terms
in general~\cite{Be96,Pi97a,Pi97b,Ma98}. This lower symmetry allows to write
down new chiral invariant terms which can compete with those coming
from the anomaly rendering the anomalous (or better, pseudo-parity
odd) amplitudes temperature dependent even at leading order.

Ref.~\cite{Pi96} makes use of a linear sigma model with constituent
quarks to study the mesonic decay amplitudes at temperatures near the
chirally symmetric phase through direct computation of the relevant
Feynman diagrams. As already noted, a non trivial temperature
dependence is found and furthermore the neutral pion decay turns out
to be suppressed in the chirally symmetric phase. This conclusion is
confirmed in~\cite{Ba96} using functional methods. Both calculations
use the imaginary time formalism to introduce the finite
temperature. A similar calculation is carried out in ~\cite{Gu97}
using the real time formalism. There it is found that the pion decay
amplitude is indeed temperature dependent although the suppression in
the chirally symmetric phase is not reproduced. In ~\cite{Ge98} the
problem is studied in full generality regarding the kinematical
conditions of the pion and photons within the real time
formulation. The analysis there indicates that the discrepancy comes
from the different kinematical configurations assumed, namely, static
photons in ~\cite{Pi96} versus on-shell photons in ~\cite{Gu97}. It is
also noteworthy that, according to~\cite{Ge98}, the vanishing of the
pionic decay amplitude in the chirally symmetric phase will presumably
be recovered when quantum pionic fluctuations are properly taken into
account since they regulate the infrared sector. Another set of
calculations of anomalous (and non anomalous) mesonic amplitudes
should be mentioned here~\cite{Al94,Pi97a,Pi97b,Ma98}. They correspond
to the low temperature limit and thus well within the phase were
chiral symmetry is broken. In this case the spirit of chiral
perturbation theory applies. These calculations make use of effective
Lagrangians of pions. In this approach, the temperature dependence
comes through pionic loop corrections (instead of quark loops as in
the constituent quark models cited above) and the pionic Lagrangian
itself is assumed to be temperature independent.

In the present work we carry out a calculation of the pseudo-parity
odd component of the effective action of fermions in two and four
dimensions in the presence of external non Abelian vector, axial and
pseudo-scalar fields on the chiral circle at finite temperature, and
at leading order in a suitable expansion. We will use the
$\zeta$-function prescription and will emphasize the chiral symmetry
preservation aspects such as the anomaly and explicit vector gauge
invariance. Some implications for chiral fermions are also presented.

\section{General considerations}\label{sec:2}

The Euclidean action describing fermions in the presence of external
bosonic fields is $\int\bar\psi\D\psi$, where $\D$ is the Dirac
operator
\begin{equation}
\D=  \thru{D}+\thru{A}\gamma_5+ MU^{\gamma_5}\,,
\end{equation}
$D_\mu= \partial_\mu+V_\mu$ is the covariant derivative,
$U^{\gamma_5}$ stand for $U$ in the subspace $\gamma_5=+1$ and
$U^{-1}$ when $\gamma_5=-1$. $M$ is a constant c-number mass term, the
constituent mass of the quarks, $V_\mu(x)$, $A_\mu(x)$ and $U(x)$ are
matrices in flavor (and color) space. $M$ is real and positive,
$V_\mu(x)$ and $A_\mu(x)$ are anti-Hermitian, and $U(x)$ is
unitary. The matrices $\gamma_\mu$ and $\gamma_5$ are Hermitian and
satisfy
\begin{equation}
\{\gamma_\mu,\gamma_\nu\}= 2\delta_{\mu\nu}\,,\quad
\gamma_5=\eta_d\gamma_0\cdots\gamma_{d-1}\,,
\end{equation}
$d=2,4$ being the space-time dimension and $\eta_2=i$, $\eta_4=1$. We
will use the imaginary time formalism to implement the finite
temperature condition, namely, through periodic boundary conditions in
the Euclidean time direction for the bosonic fields and antiperiodic
conditions for the fermionic fields with period $\beta=1/T$, $T$ being
the temperature~\cite{La87,Ka89,Be96}. We will assume a space-time
topology of the form ${\rm R}^{d-1}\times{\rm S}^1$. As a consequence
Euclidean (Lorentz) invariance is partially broken due to the
anisotropic boundary conditions.

Chiral symmetry corresponds to the transformation
\begin{equation}
\D \to 
\Omega_R^{-P_L}\Omega_L^{-P_R}\D\Omega_R^{P_R}\Omega_L^{P_L} \,,
\end{equation}
where $P_{R,L}=\frac{1}{2}(1\pm\gamma_5)$ are the projectors on the
subspaces $\gamma_5=\pm 1$ and $\Omega_{R,L}(x)$ are independent
unitary flavor matrices. In terms of the chiral fields $V^{R,L}_\mu(x)
= V_\mu\pm A_\mu$, the Dirac operator can be written as
\begin{equation}
\D= (\thru{D}^R+MU)P_R+(\thru{D}^L+MU^{-1})P_L\,,
\end{equation}
where $D^{R,L}_\mu= \partial_\mu+V^{R,L}_\mu$ are the chiral covariant
derivatives, and the chiral transformation takes the form
\begin{equation}
D^{R,L}_\mu \to \Omega_{R,L}^{-1}D^{R,L}_\mu\Omega_{R,L}\,,\quad U\to
\Omega_L^{-1} U\Omega_R\,.
\label{eq:5}
\end{equation}
Vector gauge transformations correspond to the diagonal subgroup
$\Omega_R=\Omega_L$, i.e. $\D\to \Omega^{-1}\D\Omega$. For finite
rotations, $\Omega_{R,L}=\exp(\alpha_{R,L})$ the matrices
$\alpha_{R,L}$ being anti-Hermitian. Infinitesimal vector and axial
rotations are defined by $\delta\alpha_{R,L}
=\delta\alpha_V\pm\delta\alpha_A$. In the Abelian case finite axial
transformations form a group.
 
The fermionic effective action is formally defined as
$W=-\Tr\,\log(\D)$. To renormalize the ultraviolet divergent trace we
will adopt the $\zeta$-function prescription~\cite{Se67,Ha77,Ra89},
that is, $W=-\sum_n\lambda_n^s\log\lambda_n|_{s=0}$. Here, $\lambda_n$
are the eigenvalues of $\D$ and $s=0$ is understood as an analytical
continuation from sufficiently negative $s$. This method is
mathematically well founded~\cite{Se67} and it is particularly
convenient in the context of finite temperature since it automatically
preserves vector gauge invariance under both small and large gauge
transformations~\cite{De97,Sa98}. This is because, within the
$\zeta$-function prescription, the effective action depends only on
the spectrum of the Dirac operator which is invariant under gauge
transformations. Note, however, that in general the effective action
is invariant modulo $2\pi i$, due to the multivaluation of the
logarithm.

It is convenient to introduce the concept of pseudo-parity
transformation~\cite{Wi83}, namely, $V_\mu\to V_\mu$, $M\to M$,
$A_\mu\to -A_\mu$ and $U\to U^{-1}$. Due to parity invariance, which
involves and additional $(x_0,\bfx)\to(x_0,-\bfx)$ in the fields, the
pseudo-parity odd component of the effective action is that containing
the Levi-Civita pseudo-tensor. Also, it corresponds to the imaginary
part of the effective action. As it is well-known, the real part of
the effective action may be regularized preserving chiral symmetry and
the chiral anomaly is only essential (i.e., not removable by
counterterms) in the pseudo-parity odd component~\cite{Al85}. Here we
will concentrate on this latter component, which will be denoted
$W^-$.

The consistent chiral anomaly, defined as the variation of the
effective action under an infinitesimal chiral rotation, is given by
\begin{eqnarray}
\delta W^-_{d=2} &=&
\frac{i}{\pi}\int\tr\left(F-A^2\right)\delta\alpha_A\,, \cr \delta
W^-_{d=4} &=&
-\frac{1}{12\pi^2}\int\tr\left(3F^2+F_A^2-4AFA-\{F,A^2\}-A^4
\right)\delta\alpha_A\,,
\label{eq:6}
\end{eqnarray}
where we have adopted a standard differential geometry notation:
$D=D_\mu dx_\mu$, $V=V_\mu dx_\mu$, $A=A_\mu dx_\mu$, $F=D^2=dV+V^2$,
$F_A=\{D,A\}$, the $dx_\mu$ anticommute and $dx_0dx_1\cdots dx_{d-1}=
\epsilon_{01\dots d-1}d^dx$. It is noteworthy that the anomaly depends
on the gauge fields only and not on $M$ or $U(x)$ and also that there
is no anomaly associated to purely vector gauge transformations.

The anomaly can be integrated to yield the gauged WZW
action~\cite{We71,Wi83,Ka84,Pa85,Ma85}, which in two dimensions
takes the form
\begin{equation}
\Gamma(V,A,U)=
-\frac{i}{12\pi}\int\tr(R^3) 
+\frac{i}{4\pi}\int\tr\left(V^RR-V^LL+V^RU^{-1}V^LU-V^RV^L\right)\,,
\end{equation}
where $R=U^{-1}dU$ and $L=UdU^{-1}$. In the Wess-Zumino term,
$\int\tr(R^3)$, the integral refers to a three-dimensional manifold on
the gauge group which interpolates between the original $U(x)$
configuration and a fixed configuration belonging to the same homotopy
class. The latter can be taken as a constant at zero temperature. The
finite temperature case is discussed in great detail in
Ref.~\cite{Al93}. There it is shown, in particular, that $\Gamma$ does
not depend on small variations (in the topological sense) of the
interpolating path and also that the variation of $\Gamma$ is
independent of the homotopy class representative chosen. For this
invariance it is essential that the Wess-Zumino integrand is a closed
form.

Alternatively, the same gauged WZW action can also be written
as~\cite{Al85,Hu91,Ru95,Sa96,Pi97a}
\begin{equation}
\Gamma(V,A,U)=
-\frac{i}{4\pi}\int\tr\left(\frac{1}{6}\Rc^3
-\Rc F_R-2AF_R+\frac{4}{3}A^3\right) -\hbox{p.p.c}\,,
\label{eq:8}
\end{equation}
where $\Rc=R-V^R+U^{-1}V^LU$, $F_R=D_R^2=dV_R+V_R^2$ transform
covariantly under $\Omega_R$ and are invariant under $\Omega_L$, and
p.p.c stands for pseudo-parity conjugate, i.e., $A\to -A$, $U\to
U^{-1}$ (and thus exchanging the labels $R$ with $L$ everywhere). This
version can be obtained directly by starting from the Wess-Zumino
action and applying minimal coupling, $R\to R_c$. The result is then
chiral invariant but it is no longer a closed 3-form, i.e., it has a
spurious dependence on the particular interpolating path taken. This
is cured by adding new terms involving the gauge fields and preserving
vector gauge invariance. An advantage of this version is that the
chiral breaking terms are manifestly polynomial thus yielding a
polynomial anomaly. These terms cannot be removed by local polynomial
counterterms, however, since they do not form an exact three-form by
themselves. The corresponding expression in four dimensions is
\begin{eqnarray}
\Gamma(V,A,U) &=&
-\frac{1}{48\pi^2}\int\tr\Big(\frac{1}{10}\Rc^5
+2\Rc F_R^2 -\Rc^3 F_R +\Rc F_RU^{-1}F_LU+2AF_RF_L \cr
&&+ 4AF_R^2 -8A^3F_R+\frac{16}{5}A^5\Big) -\hbox{p.p.c}\,.
\label{eq:9}
\end{eqnarray}
(The $LR$ versions are given in section~\ref{sec:6}.)

We remark that the mathematical statement is that the infinitesimal
variation of $\Gamma(V,A,U)$ equals the consistent anomaly given in
the right-hand side of eqs.~(\ref{eq:6}), provided that $V$, $A$ and
$U$ transform under chiral rotations as in
eq.~(\ref{eq:5}). Therefore, $U$ can actually stand for any field
taking values on the gauge group and transforming as
$\Omega_L^{-1}U\Omega_R$. This is true when $U$ is the pseudo-scalar
field appearing in the Dirac operator, but would apply also for a $U$
suitably constructed out of $V$ and $A$, for instance. This remark is
specially relevant when $M=0$, since then there is no pseudo-scalar
field in the Dirac operator. In any case, by construction, the gauged
WZW term $\Gamma(V,A,U)$ and the effective action $W^-(V,A,U;M,T)$
differ by chiral covariant terms only.

In order to carry out the ulterior calculation we will fix the chiral
gauge: by taking $\Omega_R=\Omega$ and $\Omega_L= U\Omega$ a new field
configuration is obtained such that $U=1$. Let us denote the new gauge
fields by $\Vr_\mu$ and $\Ar_\mu$. The field $\Omega(x)$ represents a
vector gauge freedom. It will be convenient to partially fix this
gauge by imposing stationarity of the $\Vr_0$ field. As shown in
Ref.~\cite{Sa98}, the remaining vector gauge freedom consists of two
kinds of transformations, namely, stationary gauge transformations and
discrete transformations of the form $\Omega=\exp(x_0\Lambda(\bfx))$,
where $\Lambda(\bfx)$ is any anti-Hermitian matrix commuting with
$\Vr_0(\bfx)$ and with eigenvalues of the form $2\pi in/\beta$, for
integer $n$. The quantization of $\Lambda$ ensures the preservation of
the periodic boundary conditions. In the Abelian case, the discrete
transformations are large in the topological sense.

The difference between the effective actions of the original and
rotated configurations is accounted for by the gauged WZW term, which
vanishes identically when $U(x)=1$, therefore
\begin{equation}
W^-(V,A,U;M,T) = W^-(\Vr,\Ar,1;M,T)+\Gamma(V,A,U)\,.
\label{eq:10}
\end{equation}
By construction, since the chiral gauge has been fixed, the effective
action of the $(\Vr,\Ar,\Ur=1)$ configuration defines a chiral
invariant action when re-expressed in terms of the original fields. In
the actual calculation, which necessarily truncates the exact result
to some order, chiral invariance will be maintained provided that the
calculation in terms of the rotated fields preserves vector gauge
invariance.

\section{Calculation of the effective action}
\label{sec:3}

In this section we will explain our calculation of the effective
action at finite temperature and the expansion used. The result of the
calculation is discussed in the next section. Let us introduce the
$\zeta$-function
\begin{equation}
\Omega_s(\D)=\Tr(\D^s)=\sum_n\lambda_n^s\,,
\end{equation}
which is ultraviolet finite if Re$(s)<-d$ and a meromorphic function
of $s$ with simple poles at $s=-d,-d+1,\dots,-1$ ~\cite{Se67}. Then, the
$\zeta$-function prescription is
\begin{equation}
W(\D)= -\frac{d}{ds}\Omega_s(\D)\Big|_{s=0}\,.
\end{equation}
Applying Cauchy theorem
\begin{equation}
\Omega_s= -\Tr\int_\Gamma\frac{dz}{2\pi i}\frac{z^s}{\D-z}\,,
\end{equation}
where the integration path in the complex plane starts at infinity
following for instance the negative real axis, encircles the origin
clockwise and goes back to infinity along the same
ray~\cite{Se67,Sa96}. In this regularization, the mass-like term $z$
is responsible for the explicit breaking of chiral symmetry needed to
allow for the chiral anomaly.

To deal with the trace on the space-time degrees of freedom, we apply
the Wigner transformation method~\cite{Sa96,Sa98}:
\begin{equation}
\Omega_s= -\int\frac{d^{d-1}k}{(2\pi)^{d-1}} \frac{1}{\beta}\sum_n
\int_\Gamma\frac{dz}{2\pi i}z^s\tr \langle
0|\frac{1}{\thru{p}+\D-z}|0\rangle\,.
\label{eq:13}
\end{equation}
Here, $\bfp=i\bfk$, $p_0=\omega_n$ with $\omega_n=2\pi
i(n+\frac{1}{2})/\beta$ and the sum on $n$ refers to all integers.
$\tr$ refers to flavor and Dirac spaces. $|0\rangle$ is the state of
zero momentum and zero energy in the space-time Hilbert space,
normalized as $\langle x|0\rangle=1$. In particular, this implies
$\partial_\mu|0\rangle=\langle 0|\partial_\mu=0$ whenever a derivative
inside $\D$ reaches any of the ends of the matrix element. Also,
$\langle 0|0\rangle=\int d^dx$.

In order to proceed we will work with a chiral gauge fixed
configuration, i.e., with the Dirac operator
\begin{equation}
\D=\gamma_\mu(D_\mu+A_\mu\gamma_5)+M \,, \quad \partial_0V_0=0 \,.
\end{equation}
(For simplicity, in this section we will not use a special notation
for the rotated fields. Also, for the correct interpretation of the
formulas, we remark that in this section we are not using a
differential geometry notation.) The main idea of the calculation, in
order to be able to carry out the sums and integrations indicated in
eq.~(\ref{eq:13}), is to perform a series expansion on some of the
pieces contained in the Dirac operator. Depending on the precise
choice made one can obtain perturbation theory, gradient expansions,
inverse mass expansions, etc. We want to organize the calculation so
that operators of higher dimension are also of higher order, however,
it is essential to do this preserving gauge invariance. In the Wigner
transformation formulation, the integral over $\bfk$ projects out the
contribution which is invariant under time-independent gauge
transformations whereas the sum over frequencies does the same thing
for the discrete transformations~\cite{Sa96,Sa98}. (Loosely
speaking, a time-independent gauge transformation can be compensated
by a shift in $\bfk$ and a discrete gauge transformation can be
compensated by a shift in $\omega_n$.) Therefore, gauge invariance is
preserved by expanding in powers of $\bfgamma\bfD+\thru{A}\gamma_5$
while keeping $\thru{p}+ \gamma_0D_0+M-z$ in the denominator. At
finite temperature, making a further expansion in powers of $D_0$
would break gauge invariance because $D_0$ transforms discretely under
discrete gauge transformations. The deep reason for this is that
counting powers of $\partial_0$ effectively means to study the change
of the functional under dilatations in the time coordinate, and such
dilatations are not supported by the boundary conditions.

Treating $D_0$ fully non perturbatively in the way just described is
possible since, due to our choice of chiral gauge fixing, the quantities
$p_\mu$, $\partial_0$, $V_0$ and $M-z$ appearing in the denominator
are all commuting. (See \cite{Sa98} for such a non-perturbative
treatment in three dimensions.) However, in the spirit of retaining
only lower dimensional operators, a further natural expansion is that
in powers of the quantity $\hat{D}_0$ defined by
$\hat{D}_0(X)=[D_0,X]$ which preserves gauges invariance.

Let us illustrate this kind of calculations by detailing the procedure
in the two dimensional case. Of course, due to the identity
$\gamma_\mu\gamma_5= i\epsilon_{\mu\nu}\gamma_\nu$, in this case it is
algebraically simpler to use a single non anti-Hermitian vector field
$W_\mu=V_\mu-i\epsilon_{\mu\nu}A_\mu$ in the
calculation. Nevertheless, for greater similarity with the four
dimensional case, we will keep the vector and axial fields as
independent variables .  Since the leading order terms are in
principle of the form $V_0\bfA$ and $A_0\bfV$, we have to retain terms
of first and second order in the expansion of the $\zeta$-function,
\begin{eqnarray}
\Omega_{s,1+2} &=& -\int\frac{dk}{2\pi}\frac{1}{\beta}\sum_n
\int_\Gamma\frac{dz}{2\pi i}z^s\tr
\langle 0|
\Bigg(
-\frac{1}{\gamma_0Q+\bfgamma\bfp+\mu}
(\bfgamma\bfD+\thru{A}\gamma_5)
\frac{1}{\gamma_0Q+\bfgamma\bfp+\mu}
\cr &&
+\frac{1}{\gamma_0Q+\bfgamma\bfp+\mu}
(\bfgamma\bfD+\thru{A}\gamma_5)
\frac{1}{\gamma_0Q+\bfgamma\bfp+\mu}
(\bfgamma\bfD+\thru{A}\gamma_5)
\frac{1}{\gamma_0Q+\bfgamma\bfp+\mu}
\Bigg)
|0\rangle\,,
\end{eqnarray}
where $Q=\omega_n+D_0$ and $\mu=M-z$. Since $Q$, $\bfp$ and $\mu$
commute with each other we can use the identity
\begin{equation}
\frac{1}{\gamma_0Q+\bfgamma\bfp+\mu}
=\frac{\mu-\gamma_0Q-\bfgamma\bfp}{\Delta}\,,
\quad \Delta= \mu^2-Q^2+\bfk^2\,.
\label{eq:17}
\end{equation}
The Dirac trace can then be evaluated. Retaining only the
pseudo-parity odd terms one finds
\begin{eqnarray}
\Omega^-_{s,1+2} &=& 2i\int\frac{dk}{2\pi}\frac{1}{\beta}\sum_n
\int_\Gamma\frac{dz}{2\pi i}z^s\,\mu\,\tr \langle 0|\frac{1}{\Delta}
\Bigg( \bfA Q+Q\bfA
+(3\bfk^2-\mu^2)\left(A_0\frac{1}{\Delta}\bfD+\bfD\frac{1}{\Delta}A_0\right)
\cr && -QA_0\frac{1}{\Delta}[\bfD,Q]+[\bfD,Q]\frac{1}{\Delta}A_0Q
+A_0Q\frac{1}{\Delta}\bfD Q+Q\bfD\frac{1}{\Delta}QA_0
\Bigg)\frac{1}{\Delta} |0\rangle\,.
\label{eq:18}
\end{eqnarray}

Next, all the ``naked'' $\bfD$, i.e., those which are not inside
commutators, are brought to the right of the matrix element. For
instance
\begin{equation}
\bfD\frac{1}{\Delta}=\frac{1}{\Delta}\bfD+
\frac{1}{\Delta}\{Q,[\bfD,Q]\}\frac{1}{\Delta}\,.
\label{eq:19}
\end{equation}
This produces commutator terms of the form $[\bfD,X]$ which are gauge
invariant under time-independent gauge transformations. There are also
non-covariant terms, of the form $\tr\langle 0|X\bfD|0\rangle$, which
can be replaced by $\tr\langle 0|X\bfV|0\rangle$, since
$\partial_i|0\rangle=0$. Thus at the end all appearing operators are
multiplicative in $\bfx$-space, that is, $\bfx$ becomes just a
parameter and the operators effectively act on a Hilbert space which
is the tensor product of flavor and time spaces only.

The problem now is that the sum over the variables $\bfk$, $\omega_n$
and $z$ cannot be done in a straightforward manner because they appear
in different operators which do not commute. At this point, one can
insist on obtaining a strict expansion in powers of $\bfD$ and $A_\mu$
(i.e., without further expanding in powers of
$\hat{D}_0$)~\cite{Sa98}. Instead of doing so, we will simplify the
problem by retaining only terms which are of lowest dimension in
fields and derivatives. The counting is defined as follows: $M$ will
be taken of order zero, $V_\mu$, $A_\mu$ and $\partial_\mu$ count as
first order each. This is equivalent to count the number of Lorentz
indices. The treatment of $D_0$ requires some care: the quantity $Q$
is in principle of zeroth order due to $\omega_n$, however, inside
commutators $\omega_n$ does not contribute, so $Q$ becomes $D_0$ and
counts as first order. Then (recalling that expanding in powers of
$D_0$ is forbidden) the strategy to follow is to bring all naked $Q$
(the explicit ones as well as those inside $\Delta$) together to the
left of the matrix element. There they commute and the indicated sums
and integrals can be easily carried out. In doing this we keep only
leading order terms. Moving explicit the $Q$ operators to the left can
be done systematically, generating commutator terms. On the other
hand, moving $\Delta$ operators to the left is more subtle in the
sense that it cannot be done exactly in closed form. It can only be
done up to some given order in the expansion. This comes from the
identity
\begin{equation}
X\frac{1}{\Delta}=\frac{1}{\Delta}X-2\frac{Q}{\Delta}[D_0,X]\frac{1}{\Delta}
+\frac{1}{\Delta}[D_0,[D_0,X]]\frac{1}{\Delta} \,,
\label{eq:19b}
\end{equation}
which shows that new terms with a $\Delta^{-1}$ factor at the right
are generated. Nevertheless, such terms always come with commutators
and thus they are of higher order than the original one. Therefore
repeating the commutation operation a sufficient number of times, all
naked $Q$ will end up at the left, modulo higher order terms. Once
this has been achieved, all $Q$ not at the left has been replaced by
$D_0$ and they appear inside commutators. The next thing to observe is
that in all the naked $Q$ (explicit and implicit) which are now at the
left, $D_0$ can be replaced by $V_0$ since $\langle 0|\partial_0$
vanishes and $V_0$ is stationary. So finally we have a multiplicative
operator both in $\bfx$- and $x_0$-spaces and hence $\langle 0|X| 0
\rangle$ is just $\int d^dx X$. Note that there is no difficulty of
principle in computing the expansion at any given order. In practice,
it is simpler to bring the naked $Q$ to the left and the naked $\bfD$
to the right doing the truncation at the same time.

Applying this method to eq.~(\ref{eq:18}), the pseudo-parity odd
$\zeta$-function in two dimensions becomes, at leading order,
\begin{eqnarray}
\Omega^-_{s,{\rm lead.}} &=&
2i\int\frac{dk}{2\pi}\frac{1}{\beta}\sum_n \int_\Gamma\frac{dz}{2\pi
i}z^s \mu\,\tr \langle 0|\Bigg[ 2\frac{Q}{\Delta^2}\bfA
+\left(\frac{1}{\Delta^2} + 4\frac{Q^2}{\Delta^3}\right)[\bfA,D_0] \cr
&& +\left(4\frac{k^2}{\Delta^3}-\frac{1}{\Delta^2}\right)
\left([\bfD,A_0] +2A_0\bfV\right)\Bigg] |0\rangle\,,
\end{eqnarray}
where effectively $Q=\omega_n+V_0$. In this expression, the
contributions from $[\bfD,A_0]$ and $A_0\bfV$ vanish after the
integration over $k$. The cancellation of non-covariant terms of the
form $\tr\langle 0|X\bfV|0\rangle$ is a non trivial check of the
calculation. Likewise, the contribution from $[\bfA,D_0]$ vanishes
using integration by parts and cyclic property. (The cyclic property
does not apply directly here since the operation $\langle
X\rangle=\sum_n\tr\langle 0|X|0\rangle$ is not a trace, however, it
can be shown that $\langle XY\rangle=\langle YX\rangle$ whenever $X$
is multiplicative in $\bfx$-space and $Y$ is a function of
$D_0$~\cite{Sa98}.)

The only remaining term $Q\Delta^{-2}\bfA$ presents a further
subtlety. In principle, this term would be of order one but it is
actually of second order. This is because, after summing over
frequencies, the factor $Q\Delta^{-2}$ yields an odd function of $V_0$
(recall that $\partial_0$ is no longer present here) and thus this
factor is of first order. In general, the Levi-Civita pseudo-tensor
requires the saturation of $d$ Lorentz indices in a $d$-dimensional
space-time, thus the leading order term will be of order $d$.

In four dimensions the calculation of the leading contribution is
algebraically more involved, but it can be done along the same
lines. For instance, there appear terms of the form
$\langle\Delta^{-2}[[\bfA,D_0],\bfF]\rangle$, which in principle would
count as fourth order, however, using again the cyclic property, it is
clear that the non trivial contribution from $\Delta$ starts at second
order in $V_0$, so this terms is actually of sixth order. (This was to
be expected in this and similar formally ultraviolet divergent terms,
since otherwise there would be an spurious contribution to the scale
anomaly which is absent in the pseudo-parity odd sector.)

The calculation can alternatively be done using a $(d+2)$-dimensional
formalism, which relates the pseudo-parity odd part of the effective
action in $d$ dimensions with the baryon number in $d+2$
dimensions~\cite{Ho84,Ba86,Ba89,Sa96} (for instance, the two
dimensional Wess-Zumino term $(12\pi)^{-1}\int\tr{R^3}$ is the
correctly normalized baryon number in four dimensions). Such formalism
gives directly the effective action as the integral of a $(d+1)$-form,
as in eqs.~(\ref{eq:8},\ref{eq:9}). One advantage of this procedure is
that enforcement of vector gauge invariance is sufficient to fix the
action without introducing a $\zeta$-function regularization, but
working in more space-time dimensions is also more involved. The
results found with this formalism (at least in two dimensions) are
consistent with those found in the $\zeta$-function prescription.

Let us remark that fixing the chiral gauge by $U=1$ and $\partial_0
V_0=0$ is essential to carry out the calculation. Fixing $V_0$ to be
stationary allows to set $\partial_0=0$ in the operators $Q$ moved to
the left. Eliminating $U$ allows to use the identity in
eq.~(\ref{eq:17}). An exception is the case of Abelian and stationary
fields, since in this case $U$, $V_0$ and $\partial_0$ commute and
this allows to carry out the calculation without fixing $U=1$. Such
calculation can be done most conveniently using the
$(d+2)$-dimensional formalism and in particular it checks, once more,
that the anomaly is temperature independent.

\section{The effective action at finite temperature}
\label{sec:4}

Recalling the exact formula eq.~(\ref{eq:10}), we define the leading
order of the pseudo-parity odd component of the effective action as
\begin{equation}
W^-_{\rm lead.}(V,A,U;M,T) = 
W^-_{\rm lead.}(\Vr,\Ar,1;M,T)+\Gamma(V,A,U)\,,
\label{eq:24}
\end{equation}
where $\Vr$ and $\Ar$ refer to the rotated fields. The chiral
invariant term, $W^-_{\rm lead.}(\Vr,\Ar,1;M,T)$, has been computed
with the method described in the previous section and is given by
\begin{eqnarray}
W_{2,{\rm lead., c.i.}}^- &=&
-2i\int\tr\left(\varphi^{(2)}_{1,1}\bfAr\right) \,, \label{eq:22}\\
W_{4,{\rm lead., c.i.}}^- &=& \int\tr\left(
2\varphi^{(4)}_{2,1}\{\bfAr,\bfFr\}
+\left(\frac{2}{3}\varphi^{(4)}_{2,0}
+\frac{8}{3}\varphi^{(4)}_{3,2}\right)\Ar_0\{\bfAr,\bfFr_A\}
-\frac{16}{3}\varphi^{(4)}_{3,3}\bfAr^3
\right) \,, \nonumber
\end{eqnarray}
for two and four dimensions, respectively. Here, a differential
geometry notation has been used with $\bfAr=\Ar_idx_i$,
$\bfFr=\frac{1}{2}[\Dr_i,\Dr_j]dx_idx_j$ and
$\bfFr_A=[\Dr_i,\Ar_j]dx_idx_j$, and $\Ar_0$ actually stands for
$\Ar_0dx_0$. Further, the coefficients $\varphi^{(d)}_{r,m}$ are
functions of $\Vr_0$ (and $M$ and $T$) given by
\begin{eqnarray}
\varphi^{(d)}_{r,m} &=&
\partial_s\left(2r\int\frac{d^{d-1}k}{(2\pi)^{d-1}}\frac{1}{\beta}\sum_n
\int_\Gamma\frac{dz}{2\pi
i}z^s(M-z)\frac{Q^m}{\Delta^{r+1}}\right)_{s=0}\,.
\end{eqnarray}
where $Q=\omega_n+\Vr_0$ (with $\omega_n= 2\pi
i(n+\frac{1}{2})/\beta$) and $\Delta=(M-z)^2-Q^2+\bfk^2$. Besides, a
factor $dx_0$ is implicit when $m$ is odd so that in this case
$\varphi^{(d)}_{r,m}$ is a 1-form. When $m$ is even,
$\varphi^{(d)}_{r,m}$ is an even function of $\Vr_0$ and is of zeroth
order, on the other hand for odd $m$, $\varphi^{(d)}_{r,m}$ is odd in
$\Vr_0$ and it counts as first order. As a consequence all the terms
in $W^-_{d,{\rm lead.}}$ are of dimension $d$.

The following simpler formulas are equivalent whenever they are
convergent:
\begin{eqnarray}
\varphi^{(d)}_{r,m} &=&
\int\frac{d^{d-1}k}{(2\pi)^{d-1}}\frac{1}{\beta}\sum_n
\frac{Q^m}{\Delta^r}\,,
\label{eq:25}
\end{eqnarray}
where $\Delta=M^2-Q^2+\bfk^2$ and again a factor $dx_0$ should be
added when $m$ is odd. Explicitly convergent formulas for
$\varphi^{(2)}_{1,1}$ are
\begin{eqnarray}
\varphi^{(2)}_{1,1} &=&
-\frac{1}{4}\int\frac{dk}{2\pi}\left[\tanh\left(\frac{\beta}{2}
\left(\sqrt{k^2+M^2}+\Vr_0\right) \right)
- \hbox{h.c.}
\right]dx_0 \nonumber \\
&=& -\frac{1}{2}\int\frac{dk}{2\pi}\frac{\sinh(\beta\Vr_0)}
{\cosh(\beta\sqrt{k^2+M^2})+ \cosh(\beta\Vr_0)}dx_0 \,.
\label{eq:31}
\end{eqnarray}

These functions are not all independent. The following relations are useful
\begin{eqnarray}
\varphi^{(4)}_{2,1} &=& \frac{1}{4\pi}\varphi^{(2)}_{1,1}\,, \cr
\varphi^{(4)}_{3,2} &=& -\frac{1}{4}\varphi^{(4)}_{2,0}
+M^2\varphi^{(4)}_{3,0} -\frac{1}{32\pi^2} \,, \cr
\varphi^{(4)}_{3,3} &=& -\frac{1}{4}\varphi^{(4)}_{2,1} 
+M^2\varphi^{(4)}_{3,1} \,.
\end{eqnarray}
So the four-dimensional action can also be written as
\begin{equation}
W_{4,{\rm lead., c.i.}}^- = \int\tr\left(
2\varphi^{(4)}_{2,1}\{\bfAr,\bfFr\}
+\left(\frac{8}{3}M^2\varphi^{(4)}_{3,0}
-\frac{1}{12\pi^2}\right)\Ar_0\{\bfAr,\bfFr_A\}
+\frac{16}{3}\left(\frac{1}{4}\varphi^{(4)}_{2,1}
-M^2\varphi^{(4)}_{3,1}\right)\bfAr^3 \right) \,.
\label{eq:30}
\end{equation}

In two dimensions, $W^-_{\rm lead., c.i.}$ contains only a term with
the structure $\Vr_0\bfAr$. The absence of $\Ar_0\bfVr$ can be
understood as a two dimensional peculiarity. In this case the Dirac
operator depends only on the two combinations $\Vr_0-i\Ar_1$ and
$\Vr_1+i\Ar_0$ and not on the four fields independently. Because gauge
invariance requires $\Dr_1+i\Ar_0$ to appear only inside commutators,
the possible terms of dimension two of the form $\Ar_0\bfVr$ would
have to come from $[\Dr_1+i\Ar_0,\Dr_1+i\Ar_0]$ which is identically
zero. On the other hand, in four dimensions it is not clear why
$W^-_{\rm lead., c.i.}$ contains no terms with the structure
$\Ar_0\bfVr\bfVr\bfVr$, and why no time derivative should appear.

The gauge invariance of $W_{\rm lead.,c.i.}^-$ is obvious under
time-independent gauge transformations. Under discrete gauge
transformations $\Vr_0$ transforms to $\Vr_0+\Lambda$, so its
eigenvalues are shifted by an integer multiple of $2\pi i/\beta$. Due
to the sum over frequencies, $\varphi^{(d)}_{r,m}$ are periodic
functions of $\Vr_0$ with period $2\pi i/\beta$ and the result is
invariant.

Another important test is the zero temperature limit of the result. As
noted before, soft-pion theorems imply that at zero temperature, the
only pieces present at leading order in the pseudo-parity odd sector
are those coming from the gauged WZW term, which already saturates the
anomaly. Indeed, the new contributions would be chirally invariant,
and thus gauge invariant in terms of the rotated fields (i.e., in the
chiral gauge $U=1$). The possible leading order terms consistent with
Euclidean and gauge invariance are $\langle \Fr_A\rangle$ in two
dimensions and $\langle \Fr\Fr_A\rangle$ and
$\langle\Fr_A\Ar^2\rangle$ in four dimensions, all of which vanish
identically. Of course, there are new chiral invariant terms at zero
temperature at sub-leading orders, for instance, in two dimensions,
the fourth order term is (e.g. using the formulas in Ref.~\cite{Sa96})
\begin{equation}
-\frac{i}{12\pi}\frac{1}{M^2}
\int d^2x\epsilon_{\mu\nu}
\tr((\Fr_{\mu\nu}-[\Ar_\mu,\Ar_\nu])[\Dr_\lambda,\Ar_\lambda])\,.
\end{equation}
Therefore, restoration of Euclidean invariance requires $W_{\rm
lead.,c.i.}^-$ to vanish in the zero temperature limit. In this limit
$\int\frac{d^{d-1}k}{(2\pi)^{d-1}}\frac{1}{\beta}\sum_n$ becomes
$\int\frac{d^dk}{(2\pi)^d}$. All terms containing
$\varphi^{(d)}_{r,m}$ with odd $m$ vanish: since $\omega_n$ is now a
continuous variable, it can be shifted to eliminate $\Vr_0$. The
resulting integral vanish due to antisymmetry of the integrand. This
cancellation is illustrated by eq.~(\ref{eq:31}). Likewise, the
coefficient of the term $\Ar_0\{\bfAr,\bfFr_A\}$ also vanishes at zero
temperature. This can be seen using the form in eq.~(\ref{eq:30}) and
computing $\varphi^{(4)}_{3,0}$ with eq.~(\ref{eq:25}).

\section{Anomalous amplitudes}
\label{sec:5}

In this section we will obtain the mesonic amplitudes derived from the
effective action at finite temperature. We will consider only the
Abelian and stationary case, since it allows to compare with previous
calculations in the literature~\cite{Pi96,Ba96}.

In the Abelian and stationary case, $\Vr=V$ and
$\Ar=A-\frac{1}{2}d\phi$, where $U=\exp(\phi)$, thus the chiral
invariant part of the action at leading order reads
\begin{eqnarray}
W_{2,{\rm lead., c.i.}}^- &=&
-2i\int \varphi^{(2)}_{1,1}\left(\bfA-\frac{1}{2}\bfd\phi \right) \,, 
\label{eq:123}\\
W_{4,{\rm lead., c.i.}}^- &=& \int \left(
4\varphi^{(4)}_{2,1}\bfF
+\left(\frac{16}{3}M^2\varphi^{(4)}_{3,0}-\frac{1}{6\pi^2}\right)A_0\bfF_A
\right)
\left(\bfA-\frac{1}{2}\bfd\phi\right)
 \,. \nonumber
\end{eqnarray}
In order to find the amplitudes, let us retain the leading order in an
expansion in powers of $V_0$. A simple calculation, gives
\begin{eqnarray}
\varphi^{(2)}_{1,1} &=& -\frac{1}{2\pi}(1-f) V_0 + O(V_0^3) \,, \cr
\varphi^{(4)}_{2,1} &=& -\frac{1}{8\pi^2}(1-f) V_0 + O(V_0^3) \,, 
\label{eq:133}\\
\varphi^{(4)}_{3,0} &=& \frac{1}{32\pi^2 M^2}f + O(V_0^2) \,, \nonumber
\end{eqnarray}
where we have introduced the dimensionless function 
\begin{equation}
f=  \sum_{n\in Z}\frac{\pi(\beta M)^2}
{\left((\beta M)^2+\pi^2(2n+1)^2\right)^{3/2}} \,,
\end{equation}
which has limits $f\to 1$ when $T\to 0$ and $f\to 0$ when $M\to 0$.
On the other hand, in the Abelian and stationary case the gauged WZW
term reduces to
\begin{eqnarray}
\Gamma_2(V,A,U) &=& \frac{i}{2\pi}\int V_0\bfd\phi \cr
\Gamma_4(V,A,U) &=& -\frac{1}{12\pi^2}\int\left( 3V_0\bfF+A_0\bfF_A
\right)\bfd\phi \,.
\end{eqnarray}
Therefore the leading anomalous amplitudes can be read from the
following effective actions
\begin{eqnarray}
W_2 &=&
\frac{i}{2\pi}\int V_0\left(f\bfd\phi + 2(1-f)\bfA \right) \,, \\
W_4 &=& -\frac{1}{12\pi^2}\int
\left( 3V_0\bfF +A_0\bfF_A \right)
\left(f\bfd\phi + 2(1-f)\bfA \right)
 \,. \nonumber
\end{eqnarray}
These formulas suggest that the temperature dependence of the
amplitudes are independent of the space-time dimension. In the two
cases considered, introducing a finite temperature amounts to make the
substitution $\bfd\phi\to f\bfd\phi + 2(1-f)\bfA$ in the anomalous
amplitudes at zero temperature. In particular, the term
$V_0\bfF\bfd\phi$, related to the process $\pi^0\to\gamma\gamma$, gets
an extra factor of $f$ as compared with the zero temperature
amplitude. For $M \ll T$, this factor behaves as
\begin{equation}
f=\frac{7}{4\pi^2}\zeta(3)\frac{M^2}{T^2} + O\left(\frac{M^4}{T^4}\right)\,.
\end{equation}
A result which is in agreement with those in Refs.~\cite{Pi96,Ba96}.

In the zero temperature limit $f=1$ and the amplitudes reduce to the
gauged WZW term, as they should. Thus in this case there is a tight
relation between the anomalous amplitudes and the anomaly. On the
other hand, at finite temperature the amplitudes are modified, yet the
anomaly is preserved since the terms added to the gauged WZW action
are chirally invariant.

The fact that $f$ vanishes as $M\to 0$ implies that the amplitudes
involving $\phi$ cancel in the chirally symmetric phase. In principle,
such a cancellation is to be expected on general grounds (and thus it
is a non trivial test of the calculation): since the variable $U$ no
longer appears in the Dirac operator when $M=0$, the effective action
should also be independent of $U$. More generally, the effective
action should depend analytically on the external fields $V_\mu$,
$A_\mu$ and the combination $MU$. There are, however, two
considerations which affect this conclusion. First, the argument may
be spoiled by infrared divergences. For instance, at zero temperature
the dominant anomalous term is the WZW action which depends on $U$
rather than $MU$. At finite temperature, $\omega_n$ is always
different from zero so no infrared divergences should appear in the
fermionic effective action. Second, even if the argument holds for the
exact effective action, it needs not apply when only the leading order
is retained. The truncation might introduce a spurious $\phi$
dependence at $M=0$. The situation at $M=0$ beyond the Abelian and
stationary restrictions is further studied in the next section.

Another important remark is that the fact that $f\to 0$ as $M\to 0$
does not directly imply that the amplitude for $\pi\gamma\gamma$
vanish in the chirally symmetric phase. For instance, in a linear
sigma model, the relevant piece of the action is $g\bar\psi(M+\sigma
+i\gamma_5\vec\tau\vec\pi)\psi$ hence the pion field $\pi(x)$ is
related to $\phi$ as $\pi\sim M\phi$, as a consequence the relevant
amplitude goes as $f/M V_0\bfF\bfd\pi$ and the renormalization factor
due to the temperature is $f/M$. In our calculation, in agreement with
~\cite{Pi96,Ba96}, $f=O(M^2/T^2)$ and $f/M\to 0$ as $M\to 0$. However,
in Refs.~\cite{Gu97,Ge98}, $f=O(M/T)$. Whereas $f$ still vanishes as
$M\to 0$, the $\pi\gamma\gamma$ amplitude does not vanish, presumably
due to the different kinematical conditions assumed~\cite{Ge98}.

\section{Massless fermions}
\label{sec:6}

In the case of massless fermions, the gauge fixing condition $U=1$ is
meaningless because there is no field $U$ in the Dirac operator and no
chiral rotation is necessary. Thus the calculation of the effective
action described in section~\ref{sec:3} can be applied to any gauge
field configuration $(V,A)$ provided only that the requirement
$\partial_0V_0=0$ is satisfied. In order not to unnecessarily
complicate the notation, we will use the symbols $V$ and $A$ to
denote, not the original fields, but the fields after a vector
transformation so that $V_0$ is stationary. That is, the Dirac
operator is
\begin{equation}
\D=\gamma_\mu(D_\mu+A_\mu\gamma_5)\,, \quad \partial_0V_0=0 \,,
\end{equation}
and the effective action at leading order is given by $W_{\rm lead.,
c.i.}^-$ in eqs.~(\ref{eq:22}) but using $V$ and $A$ instead of $\Vr$
and $\Ar$ and with $M=0$. The required massless functions
$\varphi^{(d)}_{r,m}$ can be computed in closed form and are given by
\begin{equation}
\varphi^{(2)}_{1,1}=-\frac{1}{2\pi}V_0\,,\quad
\varphi^{(4)}_{2,1}=-\frac{1}{8\pi^2}V_0\,,
\label{eq:40}
\end{equation}
where $V_0$ actually stands for the 1-form $V_0dx_0$. (A detailed
analysis shows that in this formula $V_0$ is to be understood modulo
$2\pi i/\beta$ so that periodicity is preserved.) Note that these
functions no longer vanish as $T\to 0$ in the massless case; the two
limits $M\to 0$ and $T\to 0$ do not commute. The leading order actions
become
\begin{eqnarray}
W^-_{2,{\rm lead.}}(V,A) &=&
\frac{i}{\pi}\int\tr\left(V_0\bfA\right) \,, 
\label{eq:34} \\
W^-_{4,{\rm lead.}}(V,A) &=&
-\frac{1}{12\pi^2}\int\tr\left(
3V_0\{\bfA,\bfF\}
+ A_0\{\bfA,\bfF_A\}
+2V_0\bfA^3
\right)
\,. \nonumber
\end{eqnarray}
Consistently with dimensional arguments, these actions are temperature
independent (except, of course, through the boundary conditions). It
should be noted that they are not really local polynomial actions due
to the gauge fixing: when expressed in terms of the original fields,
these actions are in fact non-local.

In the massive case, the correct chiral transformation of the action
was not an issue, since the chiral gauge had been fixed. In the
present case, however, the correct transformation is not evident by
construction. If these formulas are applied, not directly to $(V,A)$
but to a chirally rotated configuration $(\Vr,\Ar)$ (with $\Vr_0$
stationary) chiral symmetry would require
\begin{equation}
W^-_{\rm lead.}(V,A) = W^-_{\rm lead.}(\Vr,\Ar)+\Gamma(V,A,U)\,,
\label{eq:33}
\end{equation}
where $U$ denotes the rotation. This equality holds for the exact
effective action, what is not evident is that it should hold for its
leading order too. Note that an equivalent statement is that the
right-hand side is independent of $U$, that is, there is no spurious
dependence on $U$. The discussion of the previous section shows that
this is true in the Abelian and stationary case. (In fact, the
eqs.~(\ref{eq:133}) which give $\varphi^{(2)}_{1,1}$ and
$\varphi^{(4)}_{2,1}$ at lowest order in $V_0$, turn out to coincide
with the exact ones when $M=0$, eqs.~(\ref{eq:40}).) For Abelian but
time dependent configurations, still $\Ar=A-\frac{1}{2}d\phi$, where
$U=\exp(\phi)$, and $\Vr=V$ (there is no vector gauge transformation
involved since $\partial_0V_0=\partial_0\Vr_0=0$). Then, in two
dimensions, the right-hand side of eq.~(\ref{eq:33}) yields
\begin{equation}
W^-_{\rm lead.}(V,A) + \frac{i}{2\pi}\int\bfV d_0\phi 
\label{eq:42}
\end{equation}
(where $d_0$ stands for $\partial_0 dx_0$). This formula implies that
in general there is a spurious dependence on $U$. 

In order to consider the general non-Abelian case, note that
eq.~(\ref{eq:33}) is equivalent to say that $W^-_{\rm lead.}(V,A)$
displays the correct chiral anomaly. A detailed calculation shows that
these actions do not reproduce the correct chiral anomaly in general,
but they do so for stationary configurations. It is rather remarkable
that, even if restricted to the stationary case, these actions are
able to saturate the non-Abelian anomaly in two and four
dimensions. This is more so since the calculation of $W^-_{\rm
lead.}$, detailed in section~\ref{sec:3}, knows nothing of the anomaly
or the WZW term. From this point of view, it constitutes a non trivial
check of the calculation.

The fact that the anomaly is not always reproduced means that, in the
massless case, the truncation of the effective action to its leading
order violates chiral symmetry. This is because we have done an
expansion in powers of the operator $\bfgamma\bfD+\thru{A}\gamma_5$
which is not chiral covariant; under chiral rotations it mixes with
the other piece of the Dirac operator, $\gamma_0 D_0$. This implies
that the correct anomaly is only recovered through cancellations among
the axial variation of terms of different order. (A different matter
is directly computing the axial anomaly use our expansion. The axial
anomaly within the $\zeta$-function renormalization is given by
$\Tr(-2\alpha_A\gamma_5 \D^s)_{s=0}$ ~\cite{Ga84,Sa96} which, in the
massless case, can computed with the same technique described in
section~\ref{sec:3} and gives the correct result in closed form.)

It is interesting to compare our result in the two-dimensional case
with the exact effective action for the massless Dirac operator (the
Weyl determinant) known in closed form~\cite{Ro86}. The exact result
follows from the observation that, upon complex analytical extension
of the chiral group parameters, chiral transformations are sufficient
to bring any gauge field configuration to a space-time constant
configuration. Therefore, the exact effective action is given by two
terms. First, the gauged WZW term associated to the required
analytically extended chiral rotation and second, the effective action
corresponding to the constant configuration. This latter term will
depend on the temperature, as will also depend higher orders in our
expansion. In the Abelian case (and concentrating from now on on the
temperature independent term only) the corresponding exact effective
action for the pseudo-parity odd sector is
\begin{equation}
W^-(V,A)= \frac{i}{\pi}\int V_tA_\ell \,.
\label{eq:43}
\end{equation}
The labels $\ell$ and $t$ stand for longitudinal and transverse,
respectively, that is,
\begin{equation}
V_\ell= d\left(({\partial_\mu^2})^{-1} \partial_\nu V_\nu\right)\,,
\quad V_t = V - V_\ell \,.
\end{equation}
This action is non-local thus it cannot be removed by a local
polynomial counterterm, i.e. by a suitable choice of the
renormalization prescription. In the stationary case, $V_t=V_0$ and
$A_\ell=\bfA$, so the exact effective action coincides with that given
in eq.~(\ref{eq:34}). That is, in this particular case, the leading
order is exact (in the temperature independent sector of the effective
action). On the other hand, for non stationary configurations, an
expansion of the exact result to extract its leading order is not
well-defined due to the presence of $\partial^{-2}$ in the definition
of $V_\ell$, i.e., due to infrared divergences in the expansion.

The exact Weyl determinant is not known in four dimensions. The trick
of analytical extension of the chiral group only doubles the dimension
of the chiral group and thus it is insufficient to rotate to zero all
the components of the gauge configuration. Nevertheless, in the
Abelian case it is easy to write down a pseudo-parity odd action which
saturates the correct anomaly, namely,
\begin{equation}
W^-(V,A)=
\frac{1}{12\pi^2}\int \left( 2FAV_t-FVA_\ell-F_AAA_\ell \right) \,.
\label{eq:45}
\end{equation}
(A version of it exists for any even number of dimensions.) Unlike the
two dimensional case, this action does not reduce directly to that
given in eq.~(\ref{eq:34}) for stationary configurations. This does
not necessarily imply that they are inconsistent with each other,
since the difference are terms which are subject to infrared
ambiguities if one insists on a gradient expansion.

Of course, for massless fermions it is much more natural to work with
chiral fields. Since right and left fields are completely decoupled,
the effective action must satisfy
\begin{equation}
W^-(V,A)=W^-(V^R)-W^-(V^L)+P(V^R,V^L)\,,
\label{eq:46}
\end{equation}
where $P$ is a local polynomial introduced by the renormalization
prescription. In terms of the chiral field $v$ (say $V^R$) the
consistent anomalies take the well-known form
\begin{eqnarray}
\delta W^-_2 &=& \frac{i}{4\pi}\int\tr\left(
F-v^2
\right)\delta\alpha \,,\cr
&=& 
\frac{i}{4\pi}\int\tr\, vd\delta\alpha \,,\cr
\delta W^-_4 &=& \frac{1}{48\pi^2}\int\tr\left(
-2F^2+Fv^2+vFv+v^2F-v^4
\right)\delta\alpha \,, \cr
&=& -\frac{1}{24\pi^2}\int\tr\left(Fv-\frac{1}{2}v^3
\right)d\delta\alpha \,.
\end{eqnarray}
Where, $D=d+v$, $F=D^2=dv+v^2$ and $\delta v=[D,\delta\alpha]$. The
corresponding gauged WZW actions are
\begin{eqnarray}
\Gamma_2(v,U) &=& -\frac{i}{12\pi}\int\tr\left(R^3\right)
+\frac{i}{4\pi}\int\tr\left(vR\right)\cr
&=& -\frac{i}{12\pi}\int\tr\left(R_c^3+v^3-3(R_c+v)F \right)\,, \cr
\Gamma_4(v,U) &=& -\frac{1}{240\pi^2}\int\tr\left(R^5\right)
+\frac{1}{48\pi^2}\int\tr\left(v^3R -2FvR+v^2R^2-\frac{1}{2}vRvR+
vR^3 \right)\cr
&=& -\frac{1}{240\pi^2}\int\tr\left(R_c^5+v^5 -5(R_c^3+v^3)F+10(R_c+v)F^2
 \right)\,.
\end{eqnarray}
Here, $R=U^{-1}dU$ and $R_c=R-v$, and $U$ transforms as $U\Omega$ when
$D$ transforms as $\Omega^{-1}D\Omega$. In the $(d+1)$-dimensional
version of formulas, the polynomial term is the normalized $d+1$
Chern-Simons term and thus it yields the correct anomaly~\cite{Al85}.

The Abelian actions in Eqs.~(\ref{eq:43}) and (\ref{eq:45}) satisfy
the decoupling condition, eq.~(\ref{eq:46}), with
\begin{eqnarray}
W^-(v) &=& -\frac{i}{4\pi}\int v_\ell v\,,\quad P= \frac{i}{2\pi}\int
VA\,, \cr W^-(v) &=& \frac{1}{24\pi^2}\int v_\ell vdv \,,\quad P=
\frac{1}{6\pi^2}\int AVdV\,,
\end{eqnarray}
in two and four dimensions respectively. In each case the polynomial
$P$ is uniquely determined by vector gauge invariance. These actions
are constructed so that they saturate the anomaly, that is, they
correspond to the gauged WZW term associated to the chiral rotation
which brings the field $v$ to the gauge $v_\ell=0$. (This is achieved
by taking $U=\exp(\phi)$ where $\phi$ is defined by $v_\ell=d\phi$.)
Of course, besides the gauged WZW term the exact effective action has
another contribution (namely, the exact effective action of the purely
transverse field) but it is chiral invariant by construction and, at
least in two dimensions, it can be shown to be pseudo-parity even. The
same procedure can be followed in the $VA$ version of the theory, that
is, taking the gauged WZW term associated to axially rotate to the
gauge $A_\ell=0$. In two dimensions this procedure reproduces
eq.~(\ref{eq:43}) and thus it gives nothing new, however, in four
dimensions it yields
\begin{equation}
W^-(V,A)= -\frac{1}{12\pi^2}\int \left( 3FV+F_AA \right)A_\ell \,.
\end{equation}
This action differs from that in eq.~(\ref{eq:45}) by a chiral
invariant term of the form $\langle FV_tA_t\rangle$. Since such term
is not a local polynomial (and as far as I can see, is not identically
zero) both actions are not related by a change in the renormalization
prescription. In principle only the action in eq.~(\ref{eq:45}), which
satisfies the decoupling formula (\ref{eq:46}), could come from
integration of the fermions.

The chiral version of our leading order action corresponds to take
$V^R=v$ and $V^L=0$ in the $VA$ version, eqs.~(\ref{eq:34}). This
gives
\begin{eqnarray}
W^-_{2,{\rm lead.}}(v) &=&
\frac{i}{4\pi}\int\tr\left(v_0\bfv\right) \,,  \\
W^-_{4,{\rm lead.}}(v) &=&
\frac{1}{24\pi^2}\int\tr\left(
-v_0\{\bfv,\bfF\} +\frac{1}{2}v_0\bfv^3
\right)
\,, \nonumber
\end{eqnarray}
which refer to the gauge $\partial_0v_0=0$.

In the two dimensional case there is an important difference with the
$VA$ form of the action: due to the gauge condition,
$\partial_0v_0=0$, the only allowed infinitesimal gauge
transformations are time-independent ones. Under this restriction it
can be immediately verified that the chiral two dimensional action at
leading order always yields the correct anomaly; the configuration no
longer has to be stationary as in the $VA$ form. (In the $VA$ form of
the theory, the analogous statement holds if the formula is applied
only in the chiral gauge $\partial_0V_0= \partial_0A_0=
0$. Eq.~(\ref{eq:42}) confirms this observation since in such a gauge
$\phi$ can only be stationary and the spurious term $\bfV d_0\phi$
cancels.) This means that for any configuration, the effective action
can computed by integrating the anomaly up to a $v_0$-stationary gauge
and applying our leading order formula. The result will be independent
of the particular $v_0$-stationary gauge chosen. Unfortunately, the
action so constructed differs from the correct one by gauge invariant
terms, unless both $v_0$ and $v_1$ are stationary. What would be
needed is to ``rotate'' the original configuration to the ``gauge''
$\partial_0v_0= \partial_0v_1=0$. This can be done using an analytical
extension of the chiral group, in the same spirit as Rothe's
calculation~\cite{Ro86}.

Not unexpectedly, in four dimensions the situation is worse. The
anomaly is reproduced by $W^-_{4,{\rm lead.}}(v)$ if the configuration
is stationary, but not otherwise. Therefore it is not possible to give
a well-defined effective action in this case; the precise value would
depend on the particular $v_0$-stationary gauge taken. It would be
necessary to bring the configuration to the stationary case, but
analytical extension of the chiral group is not sufficient to do this.

\section{Summary and conclusions}

We have studied the temperature dependence of the effective action of
fermions in the presence of external bosonic fields. Due to the
complexity of the problem, only the leading order terms, that is,
those which can compete with the anomalous ones, have been retained.
However, the calculation could in principle be carried out to higher
orders as well. The temperature dependence of the anomalous
amplitudes is shown to be fully consistent with chiral symmetry
including the known temperature independence of the axial anomaly.
The known zero temperature limit, given by soft-pions theorems, is
verified. Also, for temperatures near chiral symmetry restoration,
previous results in the literature obtained with the imaginary time
formalism for static final states are reproduced. Finally, some non
trivial tests are also verified in the case of massless fermion, but
full $U$-independence of the effective action as $M\to 0$ is only
reproduced in the stationary case, due to the truncation at leading
order.

\section*{Acknowledgments}
This work is supported in part by funds provided by the Spanish DGICYT
grant no. PB95-1204 and Junta de Andaluc\'{\i}a grant no. FQM0225.


\begin{references}
\bibitem{Do74}
L. Dolan and R. Jackiw, Phys. Rev. {\bf D9}, 3320 (1974).
\bibitem{It83} H. Itoyama and A.H. Mueller, 
	Nucl. Phys. {\bf B218}, 349 (1983).
\bibitem{Re85} M. Reuter and W. Dittrich, Phys. Rev. {\bf D32}, 513 (1985).
\bibitem{Ru86} F. Ruiz Ruiz and R. \'Alvarez-Estrada,
	Phys. Lett. {\bf B180}, 153 (1986).
\bibitem{Da87} A. Das and A. Karev, Phys. Rev. {\bf D36}, 623 (1987).
\bibitem{Co88} C. Contreras and M. Loewe, Z. Phys. {\bf C40}, 253 (1988).
\bibitem{Li88} Y-L. Liu and G-J. Ni, Phys. Rev. {\bf D38}, 3840 (1988).
\bibitem{Wa90} R-T. Wang, J. Phys. {\bf A23}, 5555 (1990).
\bibitem{He90} X-G. He and G.C. Joshi, Phys. Rev. {\bf D41}, 3796 (1990).
\bibitem{Ba91} R. Baier and E. Pilon, Z. Phys. {\bf C52}, 339 (1991).
\bibitem{Ra92} A.A. Rawlinson, D. Jackson and R.J. Crewther, 
	Z. Phys. {\bf C56}, 679 (1992).
\bibitem{Sm92} A.V. Smilga, Phys. Rev. {\bf D45}, 1378 (1992).
\bibitem{Go94} A. G\'omez Nicola and R.F. \'Alvarez-Estrada,
	Int. J. Mod. Phys. {\bf A9}, 1423 (1994).
\bibitem{La87} N.P. Landsman and C.G. Van Weert,
	Phys. Rep. {\bf 145}, 141 (1987).
\bibitem{Ka89} J. Kapusta, {\em Finite Temperature Field Theory},
	Cambridge University Press (Cambridge, 1989).
\bibitem{Be96} M. Le Bellac, {\em Thermal Field Theory},
	Cambridge University Press (Cambridge, 1996).
\bibitem{Al85} L. \'Alvarez-Gaum\'e and P. Ginsparg, 
	Ann. Phys. (N.Y.) {\bf 161}, 423 (1985). Erratum, {\em ibid}
	{\bf 171}, 233 (1986).
\bibitem{Sa98} L.L. Salcedo, {\em Parity breaking in 2+1 dimensions
	and finite temperature}, hep-th/9802071.
\bibitem{Pi96} R.D. Pisarski, Phys. Rev. Lett. {\bf 76}, 3084 (1996).
\bibitem{Ad69} S.L. Adler, Phys. Rev. {\bf 177}, 2426 (1969).
\bibitem{Wi83} E. Witten, Nucl. Phys. {\bf B223}, 422 (1983).
\bibitem{We71} J. Wess and B. Zumino, Phys. Lett. {\bf B37}, 95 (1971).
\bibitem{Pi97a} R.D. Pisarski, T.L. Trueman and M.H.G. Tytgat,
	Phys. Rev. {\bf D56}, 7077 (1997).
\bibitem{Pi97b} R.D. Pisarski and M. Tytgat, Phys. Rev. Lett. 
	{\bf 78}, 3622 (1997).
\bibitem{Ma98} C. Manuel, Phys. Rev. {\bf D57}, 2871 (1998).
\bibitem{Ba96} R. Baier, M. Dirks and O. Kober, Phys. Rev. {\bf D54},
	2222 (1996).
\bibitem{Gu97} S. Gupta, S.N. Nayak, preprint TIFR/TH/97-03.
\bibitem{Ge98} F. Gelis, {\em Ambiguities in the zero momentum limit of
	the thermal $\pi^0\gamma\gamma$ triangle diagram}, preprint
	LAPTH-689/98, hep-ph/9806425.
\bibitem{Al94} R.F. \'Alvarez-Estrada, A. Dobado and A. G\'omez
	Nicola, Phys. Lett. {\bf B324}, 345 (1994).
\bibitem{Se67} R.T. Seeley, Amer. Math. Soc. Proc. Symp. Pure Math.
	{\bf 10}, 288 (1967).
\bibitem{Ha77}  S.W. Hawking, Comm. Math. Phys. {\bf 55}, 133 (1977).
\bibitem{Ra89} P. Ramond, {\em Field theory: a modern primer}, Addison Wesley
	(Frontiers in Physics 74), 1989.
\bibitem{De97} S. Deser, L. Griguolo and D. Seminara, Phys. Rev. Lett.
	{\bf 79}, 1976 (1997).
\bibitem{Ka84} \"O. Kaymak{\c c}alan, S. Rajeev and J. Schechter,
	Phys. Rev. {\bf D30}, 594 (1984).
\bibitem{Pa85} N.K. Pak and P. Rossi, Nucl. Phys. {\bf B250}, 279 (1985).
\bibitem{Ma85} J.L. Ma\~nes, Nucl. Phys. {\bf B250}, 369 (1985).
\bibitem{Al93} R.F. \'Alvarez-Estrada, A. Dobado and A. G\'omez
	Nicola, Phys. Lett. {\bf B319}, 238 (1993).
\bibitem{Hu91} C.M. Hull and B. Spence, Mod. Phys. Lett. {\bf A6}, 969 (1991).
\bibitem{Ru95} E. Ruiz Arriola and L.L. Salcedo, Nucl. Phys. 
	{\bf A590}, 703 (1995).
\bibitem{Sa96} L.L. Salcedo and E. Ruiz Arriola, Ann. Phys. (N.Y.)
	{\bf 250}, 1 (1996).
\bibitem{Ho84} E. D'Hoker and E. Farhi, Nucl. Phys. {\bf B248}, 59 (1984).
\bibitem{Ba86} R. Ball and H. Osborn, Nucl. Phys. {\bf B263}, 245 (1986).
\bibitem{Ba89} R. Ball, Phys. Rep. {\bf 182}, 1 (1989).
\bibitem{Ga84} R.E. Gamboa Sarav\'{\i}, M.A. Muschietti, F.A. Schaposnik
	and J.E. Solomin, Ann. Phys. (N.Y.) {\bf 157}, 360 (1984).
\bibitem{Ro86} K.D. Rothe, Nucl. Phys. {\bf B269}, 269 (1986)
\end{references}
\end{document}